\documentclass[reprint,amsmath,amssymb,aps,prd,nofootinbib,showkeys]{revtex4-2}
\usepackage[utf8]{inputenc} 
\usepackage[T1]{fontenc} 
\usepackage{graphicx}
\usepackage{dcolumn}
\usepackage{bm}
\usepackage{aas_macros}
\usepackage{amsfonts} 
\usepackage{microtype}
\usepackage{booktabs}
\usepackage{tabularx}
\usepackage{nicefrac} 
\usepackage{url}
\usepackage{orcidlink}
\usepackage{natbib}
\usepackage{hyperref}
\usepackage{nameref}

\setcitestyle{authoryear,round}
\begin{document} 

% \title{Homogeneity scale in the local Universe: \\
% Model-independent estimate from S-PLUS iDR6 blue galaxies}
\title{Homogeneity scale in the local Universe:%
\texorpdfstring{\\}{ }%
Model-independent estimate from S-PLUS iDR6 blue galaxies}

\author{Camila Franco\orcidlink{0000-0002-6320-425X}}
\email{camilafranco@on.br}
 \affiliation{Observatório Nacional, Rua General José Cristino, 77, São Cristóvão, 20921-400, Rio de Janeiro, RJ, Brazil}

\author{Felipe Avila\orcidlink{0000-0002-0562-2541}}
\email{felipeavila@on.br}
\affiliation{Observatório Nacional, Rua General José Cristino, 77, São Cristóvão, 20921-400, Rio de Janeiro, RJ, Brazil}

\author{Armando Bernui~\orcidlink{0000-0003-3034-0762}}\email{bernui@on.br}
\affiliation{Observatório Nacional, Rua General José Cristino, 77, São Cristóvão, 20921-400, Rio de Janeiro, RJ, Brazil}

\author{Ulisses Ribeiro}
\affiliation{Centro Brasileiro de Pesquisas F\'isicas, Rua Dr. Xavier Sigaud 150, 22290-180 Rio de Janeiro, RJ, Brazil}

\author{Clécio R. Bom~\orcidlink{0000-0003-4383-2969}}\email{debom@cbpf.br}
\affiliation{Centro Brasileiro de Pesquisas F\'isicas, Rua Dr. Xavier Sigaud 150, 22290-180 Rio de Janeiro, RJ, Brazil}

\author{Arianna Cortesi}
\affiliation{Instituto de F\'{i}sica, Universidade Federal do Rio de Janeiro, 21941-972, Rio de Janeiro, Brazil}\affiliation{Observat\'{o}rio do Valongo, Ladeira Pedro Antonio, 43, 20080-090, Rio de Janeiro, Brazil}

\author{E. Telles}
\affiliation{Observatório Nacional, Rua General José Cristino, 77, São Cristóvão, 20921-400, Rio de Janeiro,  RJ, Brazil}

\author{W. Schoenell}
% \affiliation{GMTO Corporation 465 N. Halstead Street, Suite 250 Pasadena, CA 91107, USA}
\affiliation{The Observatories of the Carnegie Institution for Science, 813 Santa Barbara St, Pasadena, CA 91101, USA}

\author{T. Ribeiro}
\affiliation{Rubin Observatory Project Office, 950 N. Cherry Ave., Tucson, AZ 85719, USA}

\author{A. Kanaan}
\affiliation{Departamento de F\'isica, \\ Universidade Federal de Santa Catarina, Florian\'opolis, SC, 88040-900, Brazil}

\author{C.~Mendes de Oliveira}
\affiliation{Universidade de S\~ao Paulo, IAG, Rua do Matão 1225, São
Paulo, SP, Brazil}

\date{\today}

%------------------------------
\begin{abstract}
We present a model-independent estimate of the angular homogeneity scale in the local Universe by analysing data from the Southern Photometric Local Universe Survey (S-PLUS).
Two complementary estimators are employed: (i) a parametric approach fitting the power-law of the two-point angular correlation function, which yields the homogeneity scale $\theta_H = 8.94_{-3.38}^{+1.20}\;{\rm deg}$; and 
(ii) a nonparametric fractal correlation dimension method, computing $\mathcal{D}_2(\theta)$ directly from the correlation function, which results in $\theta_H = 7.91_{-2.94}^{+5.50}\;{\rm deg}$. 
We apply both methodologies to a set of mock catalogs, generated with the 
Generator for Large Scale Structure algorithm, and find that both estimates are well within $1 \sigma$ of their respective median values. 
The transition scale to homogeneity, according to the $\Lambda$ cold dark matter (CDM) model, is defined for matter, i.e., $b = 1$. 
Measurements of this scale with observational data clearly depend on the cosmic tracer analyzed, and a calibration is necessary. 
Our study with blue galaxies, with bias $b \simeq 1$, provides a suitable estimate for comparison. 
Indeed, the results obtained in both approaches 
are compared with the value expected in the $\Lambda$CDM model, obtaining a good concordance.
\end{abstract}


\maketitle

%------------------------------
\section{Introduction}~\label{sec:intro}
The cosmological principle (CP), one of the pillars of 
the standard cosmological model, postulates that on sufficiently large scales 
--and at any epoch-- the universe should appear homogeneous and isotropic on average. 
Empirically, probing the validity of the CP with diverse cosmic tracers and at different epochs of the Universe's evolution, is of paramount importance for the standard model, since it underpins the theoretical framework of the $\Lambda$ cold dark matter (CDM) and its predictions for structure formation. 

Isotropy and homogeneity properties have been extensively tested using different tracers and cosmological observables~\citep{Bolejko2009, Maartens2011}, such as quasars~\citep{Secrest2021,Goncalves2018b,Fujii2022}, 
cosmic microwave background (CMB)~\citep{Aluri2012, Khan2022b, Novaes2016, Planck2020-iso}, CMB lensing potential map~\citep{Marques2018}, 
gravitational waves~\citep{Galloni2022}, $\gamma$-ray bursts~\citep{Jakub2019, Lopes2025}, 
galaxies~\citep{Labini2010, Scrimgeour2012, Courtois13, Alonso2014, Appleby2014, Goncalves2018, Avila2019, Sarkar2019}, 
supernovae Ia~\citep{Chang2019, Perivolaropoulos2023, Lopes2024, Mokeddem2026}, 
and HI extragalactic sources~\citep{Avila2023, Franco2024, Wu2025, Bizarria2025}, 
among others.\footnote{
Notice that the literature reports analyses of cosmic tracers indicate, with different levels of statistical significance, a dipolar anisotropy (see, e.g.,~\cite{Aluri2022, Kester2024}).} 
On the other hand, the homogeneity test presents observational challenges, especially when it comes to the use of photometric surveys, where the loss of radial information due to errors in redshift determination requires alternative approaches, such as purely angular analyses~\citep{Alonso2014}.

Traditionally, some studies use counts on spheres~\citep{Scrimgeour2012} and on spherical caps~\citep{Laurent2016} to determine the transition scale to homogeneity, $R_{H}$ and $\theta_{H}$ respectively, beyond which the distribution of matter tends asymptotically to the homogeneous regime. 
Studies on three-dimensional cosmic homogeneity have been carried out recently~\citep{Avila2022,Ntelis2017,RG2021}; however, methods based on physical distances rely on cosmological assumptions to convert redshifts into distances, introducing potential biases~\citep{Clarkson2012}. 
One way to get around this issue is through model-independent analyses, which employ only angular coordinates~\citep{Alonso2014,Avila2019,Avila2018,Goncalves2018,Fanha2026}. 

%explore the angular cosmic homogeneity
Thus, in this work we search for the angular scale of transition to cosmic homogeneity using photometric data from the Southern Photometric local Universe Survey (S-PLUS). 
We will apply the scaled counts-in-caps estimator and the fractal correlation dimension to determine $\theta_{H}$ in a model-independent manner.

This work is organized as follows: Sec.~\ref{sec:splus} describes the observational data used 
in our analyses. In Sec.~\ref{sec:fractal}, we outline the theoretical framework of the fractality. 
The log-normal simulations and the method to obtain them are presented in Sec.~\ref{sec:mocks}. 
The results are discussed in Sec.~\ref{sec:results}. 
Finally, robustness tests are detailed in Sec.~\ref{sec:robustness}, and we summarize our conclusions in Sec.~\ref{sec:conclusions}.

%------------------------------
\section{The Southern Photometric Local Universe Survey}
\label{sec:splus}
The Southern Photometric Local Universe Survey~\citep[S-PLUS; ][]{Mendes2019} is an ongoing photometric survey designed to cover $\sim 9000$ deg$^{2}$ across $12$ filters using the T80-South telescope, located at the Cerro Tololo Observatory (Chile). 
Currently, the survey is in its public data release 5 (DR5), covering about $4,600$ deg${^2}$~\citep{Herpich2024, Schwarz2025}, and in its internal data release 6 (iDR6).

The role of astrophysical parameters of red and blue galaxies, such as color and luminosity, in the process of matter clustering is being studied and revealed~\citep{Zehavi05, Croton07, Ross14, Mohammad18}. 
Blue galaxies are, in their majority, late-type galaxies with significant star formation, unlikely to be found in high-density regions~\citep{Dressler, Gerke07}, 
a feature that is reflected in clustering statistics as the two-point correlation function, where, on small scales, red galaxies of any luminosity are more clustered than blue galaxies of any luminosity~\citep{Zehavi05}, meaning that blue galaxies exhibit weaker clustering features compared to red galaxies. In fact, blue galaxies are found in low-density regions where they exhibit reduced nonlinear clustering effects, making the class of blue galaxies the suitable cosmic tracer for our analyses~\citep{Gerke07, Mohammad18, deCarvalho2021, Avila2024}. Moreover, blue galaxies account for nearly $60\%$ of galaxies in the local Universe~\citep{Bamford2009}, being, therefore, also more numerous in relation to 
red galaxies. 

For these reasons we follow the procedure described in~\cite{Ribeiro2025} and select for our study a sample of blue galaxies from the S-PLUS iDR6. 
Specifically, we use the following criteria: 
\begin{itemize}
\item ($r$\_\textsc{auto} $>= 14$) \textsc{and} ($r$\_\textsc{auto} $\le 18$);
\item \textsc{PROB\_GAL\_GAIA} $> 0.9$;
\item \textsc{FWHM\_n} $\ge 1.5$;
\item \textsc{SEX\_FLAGS\_r} $\le 2$;
\item \textsc{CLASS\_STAR} $\le 0.2$;
\item \textsc{CLASS\_GAIA} $= 2$;
\item \textsc{odds} $\ge 0.4$,
\end{itemize}
\noindent
where \textsc{PROB\_GAL\_GAIA} and \textsc{CLASS\_GAIA} correspond to the star-galaxy probabilities derived from the model discussed by \cite{nakazono2021discovery}. 
To increase the classification reliability, cuts in magnitude were also applied in order to avoid very large and faint objects which are often misclassified \citep{nakazono2021discovery}. Additionally, cuts in the odds parameter were performed to improve the redshift quality \citep{lima2022photometric}. The remaining selections are direct outputs from SExtractor source extraction pipeline employed in S-PLUS; these parameters serve as complementary quality indicators and provide additional support for the star–galaxy classification model.

We also applied the same color-color selection adopted by the authors of Ref.~\cite{Avila2019} to isolate blue galaxies, that is, 
\begin{itemize}
\item $(0 < g - r < 0.6)\,\textsc{and}\,(0 < u - r < 2.0)$.
\end{itemize}

\noindent
To perform this selection, objects with large photometric uncertainties were excluded and K-corrections were applied following \cite{bom2024extended}. 

\begin{figure}[!ht]
\begin{minipage}[b]{\linewidth}
\centering
\includegraphics[width=0.9\textwidth]{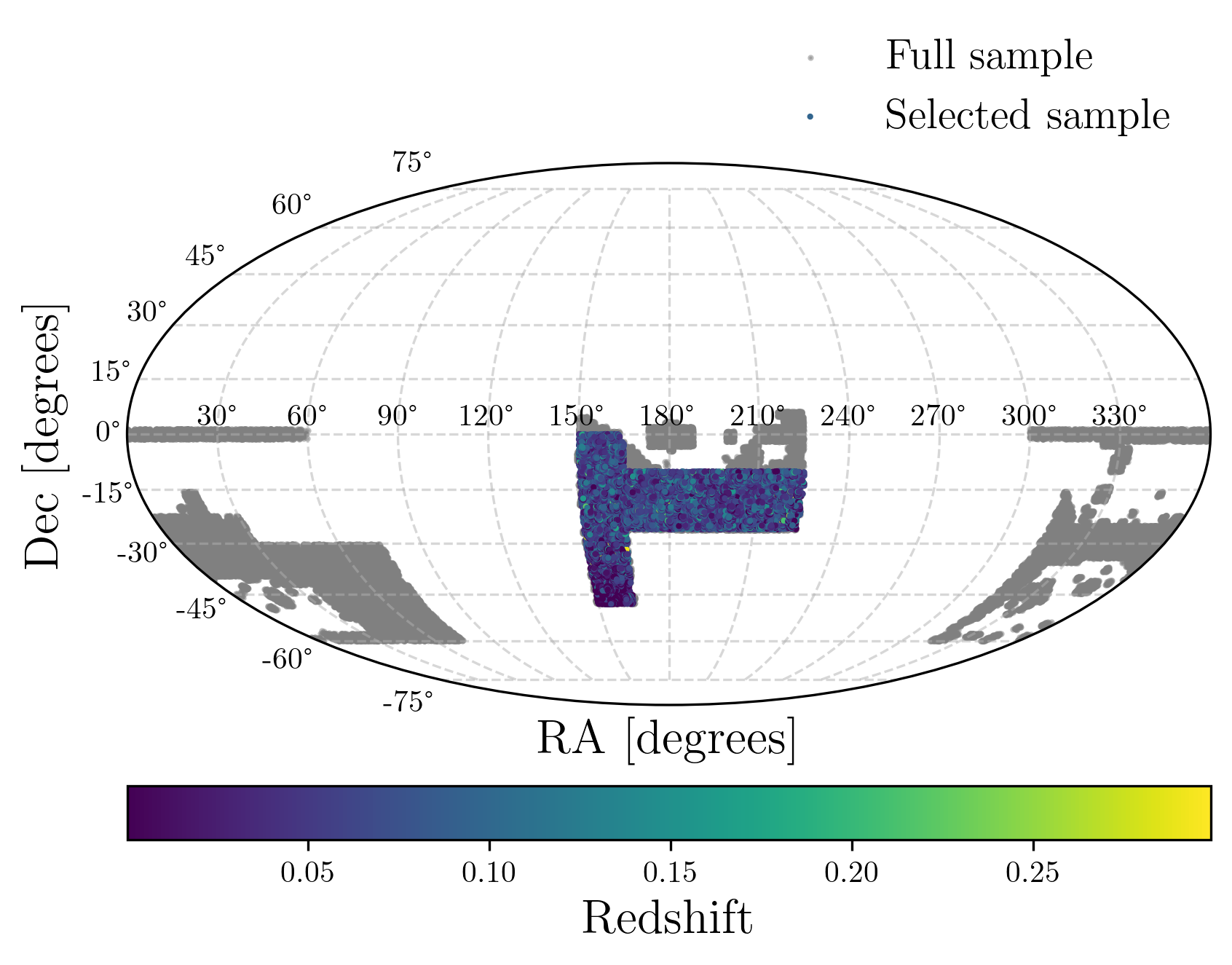}
\includegraphics[width=0.9\textwidth]{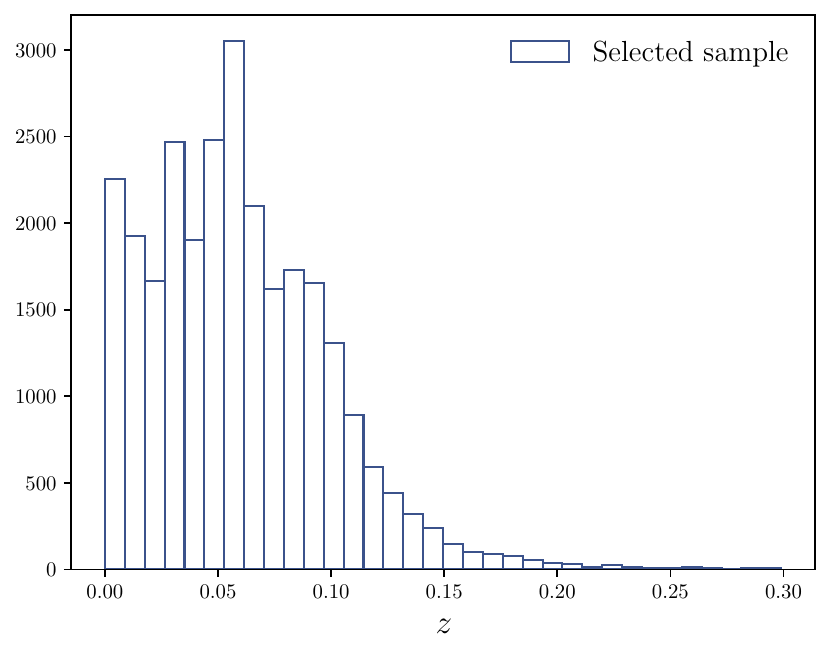}
\end{minipage}
\caption{Distribution of the S-PLUS blue galaxies corresponding to iDR6. 
Upper panel: Sky coverage of the S-PLUS (gray) and the selected sample (blue) in equatorial coordinates. The color scale represents the redshift range. 
Lower panel: Redshift distribution of the sample selected for our analyses, containing $27,289$ blue galaxies.}
\label{fig:splus}
\end{figure}

We focus on a subsample within angular coordinates 
$150^{\circ}\leq{\rm RA}\leq226^{\circ}$ and 
$-48^{\circ}\leq{\rm Dec}\leq0^{\circ}$, that is, covering an area of $1,696\,\text{deg}^2$, 
and in the redshift range $0 \leq z \leq 0.3$ (these are photometric redshifts, measured, processed, and derived by the S-PLUS team, released in the iDR6 data).
The selected sample contains $27,289$ blue galaxies.
The choice is driven because it is a contiguous completely observed sky patch, with suitable number density for the purposes of our analysis. 
The angular distribution of the full S-PLUS iDR6 data and of the selected sample for analyses are shown in the upper panel of Fig.~\ref{fig:splus}, while the lower panel presents the redshift distribution of the sample 
selected for our analyses.

%%-----------------------------------------------------------
\section{Fractality and Homogeneity estimators}
\label{sec:fractal}
The fractality concept plays an important role in analyses of cosmic structures distribution. 
A fractal ensemble is characterized by its fractal dimension, $\mathcal{D}$, which quantifies how the number of objects contained in a region of radius $r$ scales with the size of this region. 
One expects that, at sufficiently large scales, an homogeneous and isotropic universe  has a count number of objects proportional to the volume. 
On the other hand, at small scales, the distribution of galaxies exhibits fractal properties, reflecting the hierarchical structure of the cosmos (see, e.g., Refs.~\cite{Pietronero1987, 
Pietronero1992, Alonso2014, Ntelis2017, Avila2018, Bagla2008, Yadav2010} for more information about fractality). 

In this study we do not assume a fiducial cosmology because we only use angular 
distances, in this sense our analyses provide model-independent estimates of 
the angular scale of homogeneity $\theta_{H}$. 
In fact, we will analyze the data projected on the celestial sphere using the 
counts-in-spherical-caps estimator $N(< \theta)$. 
This function is defined as the mean number of points contained in a spherical cap of 
radius $\theta$, centered in some point of reference. 
However, due to the introduction of bias by the survey geometry, such as 
boundary effects, and possible observational incompleteness of the survey, 
it is not suitable to directly use $N(< \theta)$. 
Aiming to reduce these effects it is introduced the estimator 
scaled counts-in-spherical-caps, $\mathcal{N}(< \theta)$, originally defined as~\citep{Scrimgeour2012,Alonso2014} 
\begin{equation}\label{eq:n2_2d}
\mathcal{N}(<\theta) \equiv \frac{N_{ {\rm gal}}(<\theta)}{N_{{\rm rand}}(< \theta)} \,, 
\end{equation}
where $N_{{\rm gal}}(< r)$ is the average counting estimated for spherical caps centered at each cosmic object of the data catalog, and $N_{{\rm rand}}(< r)$ is the same quantity calculated upon a random homogeneous sample, using as centers the coordinate positions of the data catalog sources. 
Notice that the random catalog has the same geometry and completeness of the observational data, but with a significant higher density;  
for our analyses, we constructed a random catalog ten times denser than the sample of blue galaxies in the study\footnote{See Appendix~\ref{app:random} for a discussion about the robustness of this choice.}.

The fractal dimension for data projected on the celestial sphere is defined 
as~\citep{Goncalves2018} 
\begin{equation}\label{eq:d2_2d}
\mathcal{D}_{2}(\theta) \equiv \frac{d\ln{\mathcal{N}(<\theta)}}{d\ln{\theta}} + \frac{\theta\sin{\theta}}{1 - \cos{\theta}} \,. 
\end{equation}
Note that the angular scale of homogeneity, $\theta_{H}$, is obtained when 
$\mathcal{D}_{2}(\theta) \rightarrow \theta\sin{\theta}/(1 - \cos{\theta})$. 
%, where $\theta \rightarrow \theta_{H}$. 
%, where $\theta_{H}$ is the angular homogeneity scale 
More precisely, one finds $\theta_H$ by adopting the $1\%$ criterion described by Ref.~\cite{Scrimgeour2012}. 
According to this, the homogeneity scale is reached when the fractal 
dimension $\mathcal{D}_{2}$ attains at $\theta = \theta_H$ the $99\%$ of its 
limiting value 
\begin{equation}\label{eq:1percent}
\mathcal{D}_{2}(\theta_{H}) = 0.99\left[\frac{\theta\sin{\theta}}{1 - \cos{\theta}}\right]_{\theta=\theta_{H}} \,.
\end{equation}
%Although it may seem to be an arbitrary criterion, it was defined and tested 
%under rigorous arguments and,
%providing the best answer when tested with simulations. 
This criterion, first defined in~\cite{Scrimgeour2012}, 
proved to be the most suitable. 
In fact, simulations constructed with a specific homogeneity scale were analyzed, and such scale was recovered with the best accuracy using this criterion.
In addition, it has an interesting feature of being survey independent, enabling it to be used in comparisons across different measurements and cosmological models.

%%-----------------------------------------------------------------------
\subsection{Landy-Szalay methodology}\label{sec:ls}

In this section we present the first approach, termed the Landy-Szalay (LS) methodology, to find $\theta_{H}$. 
This approach uses the two-point angular correlation function (2PACF), $\omega(\theta)$, to define the scaled counts-in-caps 
$\mathcal{N}(<\theta)$. 
For this scope, we adopt the LS estimator~\citep{LS93} 
\begin{equation}\label{eq:tpacf-ls}
\omega(\theta) = \frac{DD(\theta) - 2DR(\theta) + RR(\theta)}{RR(\theta)}\,,
\end{equation}
where $DD(\theta)$ is the number of data-data pairs, $RR(\theta)$ is the 
number of random-random pairs, and $DR(\theta)$ is the number of data-random 
pairs, all within an angular bin centered at $\theta$. 
These quantities are normalized by the total number of possible pairs in each case, that is 
\begin{align}
DD(\theta) = \frac{2\, dd(\theta)}{n_{g}(n_{g} - 1)}\,,\\
\notag\\
RR(\theta) = \frac{2\, rr(\theta)}{n_{r}(n_{r} - 1)}\,,\\
\notag\\
DR(\theta) = \frac{dr(\theta)}{n_{g} n_{r}} \,,
\end{align}
with $n_{g}$ ($n_{r}$) being the number of galaxies in the data (random) catalog, and \,$dd$, $rr$, $dr$\, are the raw pair-counts for each case. 
This normalization accounts for the different sample sizes of the data and random catalogs. 

Then, $\mathcal{N}(<\theta)$ is defined considering the cumulative 
scaled counts-in-caps~\citep{Avila2018,Alonso2014}, 
\begin{equation}\label{eq:count}
\mathcal{N}(<\theta) \equiv 1 + \bar{\omega}(\theta) \,,
\end{equation}
where 
\begin{equation}\label{eq:wb}
\bar{\omega}(\theta) \equiv \frac{1}{1 - \cos{\theta}} \int_{0}^{\theta} \omega(\theta')\sin \theta' d\theta' 
\end{equation}
is the average angular correlation function for a given spherical cap with angular size $\theta$.

After calculating the 2PACF given by Eq.~(\ref{eq:tpacf-ls}), we fit the 2PACF 
by a power-law function 
\begin{equation}\label{eq:2pacf-power}
\omega(\theta) = \left(\frac{\theta}{\theta_{0}}\right)^{-\beta} \,,
\end{equation}
where the parameters $\theta_0$ and $\beta$ are related to the transition scale between linear and nonlinear regimes, and the slope of the correlation, respectively~\citep{Peebles93, Connolly02, Marques20, Coil13, Kurki-Suonio, Totsuji69, Franco2025b}. 
Then we compute $\mathcal{N}(<\theta)$ using Eq.~(\ref{eq:count}), 
and $\mathcal{D}_{2}(\theta)$ using Eq.~(\ref{eq:d2_2d}).

%As for the previous method, we 
For this analysis, we compute the 2PACF by applying $24$ logarithmically spaced bins within the angular range 
$[\theta_{\text{min}}, \theta_{\text{max}}] = [0.5^{\circ}, 20^{\circ}]$, as can be seen in Fig.~\ref{fig:tpacf_ls}.
This interval was chosen as a compromise between minimizing small-scale nonlinear effects and retaining sufficiently large angular separations to probe the transition to homogeneity, and, at the same time, avoids scales that are too close to the survey footprint size (where boundary effects become more important). 
The parameters $\theta_{0}$ and $\beta$ are determined by fitting Eq.~\eqref{eq:2pacf-power} using the \textsc{CurveFit}\footnote{\url{https://docs.scipy.org/doc/scipy/reference/generated/scipy.optimize.curve_fit.html}}~\citep{scipy} package.
Then, we use Eq.~\eqref{eq:d2_2d} to calculate the correlation dimension 
$\mathcal{D}_{2}(\theta)$ by computing the logarithmic derivative of Eq.~\eqref{eq:count}, where $\mathcal{N}(\theta)$ itself is constructed using the fitted power-law form of Eq.~\eqref{eq:2pacf-power}.

%%-----------------------------------------------------------------------
\subsection{Angular fractal correlation dimension}\label{sec:d2_2d}

In this section we present the second approach, termed the 
angular fractal correlation dimension (AFCD) methodology, to find $\theta_{H}$. 
The study of the angular scale of homogeneity has the advantage of eliminating the necessity to adopt a fiducial model to convert redshifts into cosmological distances. 
That way, the two-dimensional version of the fractal dimension, $\mathcal{D}_{2}(\theta)$, is obtained through considerations analogous 
to the three-dimensional case, substituting spheres by spherical caps in 
the sky.

Using the definition of Eq.~\eqref{eq:count}, Eq.~\eqref{eq:d2_2d} becomes
\begin{equation}\label{eq:D2_wb}
\mathcal{D}_{2}(\theta) = \frac{\theta}{1+\bar{\omega}(\theta)}\frac{d\bar{\omega}(\theta)}{d\theta} + \frac{\theta\sin \theta}{1-\cos \theta}\,.
\end{equation}
Taking the derivative of Eq.~(\ref{eq:wb}) with respect to $\theta$, 
including it in Eq.~(\ref{eq:D2_wb}), and simplifying terms, 
one obtains (details of the deduction of this equation are presented in Appendix~\ref{appendixA}) 
\begin{equation}\label{eq:d2_omegas}
\mathcal{D}_{2}(\theta) = \frac{\theta \sin \theta}{1 - 
\cos \theta}\left[\frac{1 + \omega(\theta)}{1 + \bar{\omega}(\theta)}\right] \,.
\end{equation}
To compute the 2PACF, we used the public code \textsc{treecorr}\footnote{\url{https://rmjarvis.github.io/TreeCorr/_build/html/index.html}}~\citep{Jarvis2015}, applying $24$ logarithmically spaced bins within the angular range $[\theta_{\text{min}}, \theta_{\text{max}}] = [0.5^{\circ}, 20^{\circ}]$.
A random catalog was constructed, maintaining the same angular footprint as the original region but with a uniform distribution of points and a number density $10$ times greater than that of the observed dataset.

%------------------------------
\section{GLASS simulations}
\label{sec:mocks}

In our analyses we use mock realizations produced by the \textsc{GLASS} (Generator for Large Scale Structure;~\cite{Tessore23})\footnote{\url{https://glass.readthedocs.io/stable/index.html}} numerical code. 
Codes that generate log-normal simulations can realistically reproduce the statistical properties of large observational surveys and their covariance matrix~\citep{Lippich19,Blot19,Colavincenzo19}. 
The GLASS 
%(Generator for Large Scale Structure)~\footnote{\url{https://glass.readthedocs.io/stable/index.html}}~\citep{Tessore23} 
code is particularly well suited for this work. 
It achieves these objectives by generating matter density fields in a series of spherical shells nested around the observer, effectively discretizing the past light cone, from the present day up to high redshift. 
GLASS uses the linear matter power spectrum from a given fiducial cosmology as its primary input to generate the matter density fields. 
This method allows for the efficient production of a large number of mock realizations that accurately capture the two-point statistics. 
A key feature of GLASS is its use of a log-normal model to translate the initial Gaussian density field into a more realistic, nonlinear distribution of matter.

The simulation setup defines the comoving radial resolution or redshift grid, $z_{\rm grid}$, the choice of radial window function, $W(z)$, the number of correlated redshift shells, $n_c$, the galaxy bias, $b$, the mask resolution, $n_{\rm side}$, and the maximum multipole, $l_{\rm max}$ employed to evaluate the angular power spectrum. 
It further specifies the adopted cosmological parameters, the nonlinear matter power spectrum, and the photometric redshift error $\sigma_0$ at $z=0$. 
To accurately reproduce the properties of our data, the code takes as input both 
the survey mask, and the distribution of galaxies per redshift interval and per square degree. 
Table \ref{tab:table_mock} lists the set of input values used for generating the 
$1,000$ log-normal mock catalogs for our analyses. 

\begin{table}[!ht]
\caption{Survey configuration and cosmological parameters used to generate the set of $1,000$ mock catalogs. 
We use a top-hat window function.}
\centering
\begin{tabular}{c|c}
\hline
\hline
Survey configuration & Cosmological parameters \\ 
\hline
$z_{\rm grid}=[0.0-0.3,0.05]$ & $\Omega_{c}h^{2}= 0.1202$ \\  
$n_c=6$ & $\Sigma m_{\nu}=0.06$ eV \\
$b=1.0$ & $n_{s}=0.9649$ \\
$n_{\rm nside}=2048$ & $\ln(10^{10} A_{s})=3.045$ \\
$l_{\rm max}=2048$ & $\Omega_{b}h^{2}=0.02236$ \\
$\sigma_0=0.03$ & $h=0.67021$ \\ \hline                          
\end{tabular}
\label{tab:table_mock}
\end{table}

%%-------------------------------------
\section{Results}
\label{sec:results}

In this section, we present the main results of our analyses regarding the 
estimation of the angular scale of homogeneity from clustering measurements, 
results obtained applying the two methodologies described above, 
namely, the analyses using the LS and AFCD methodologies.

%%-------------------------------------
\subsection{Results with the LS methodology}\label{sec:ls_result}

\begin{figure}[!ht]
\centering
\includegraphics[width=0.9\linewidth]{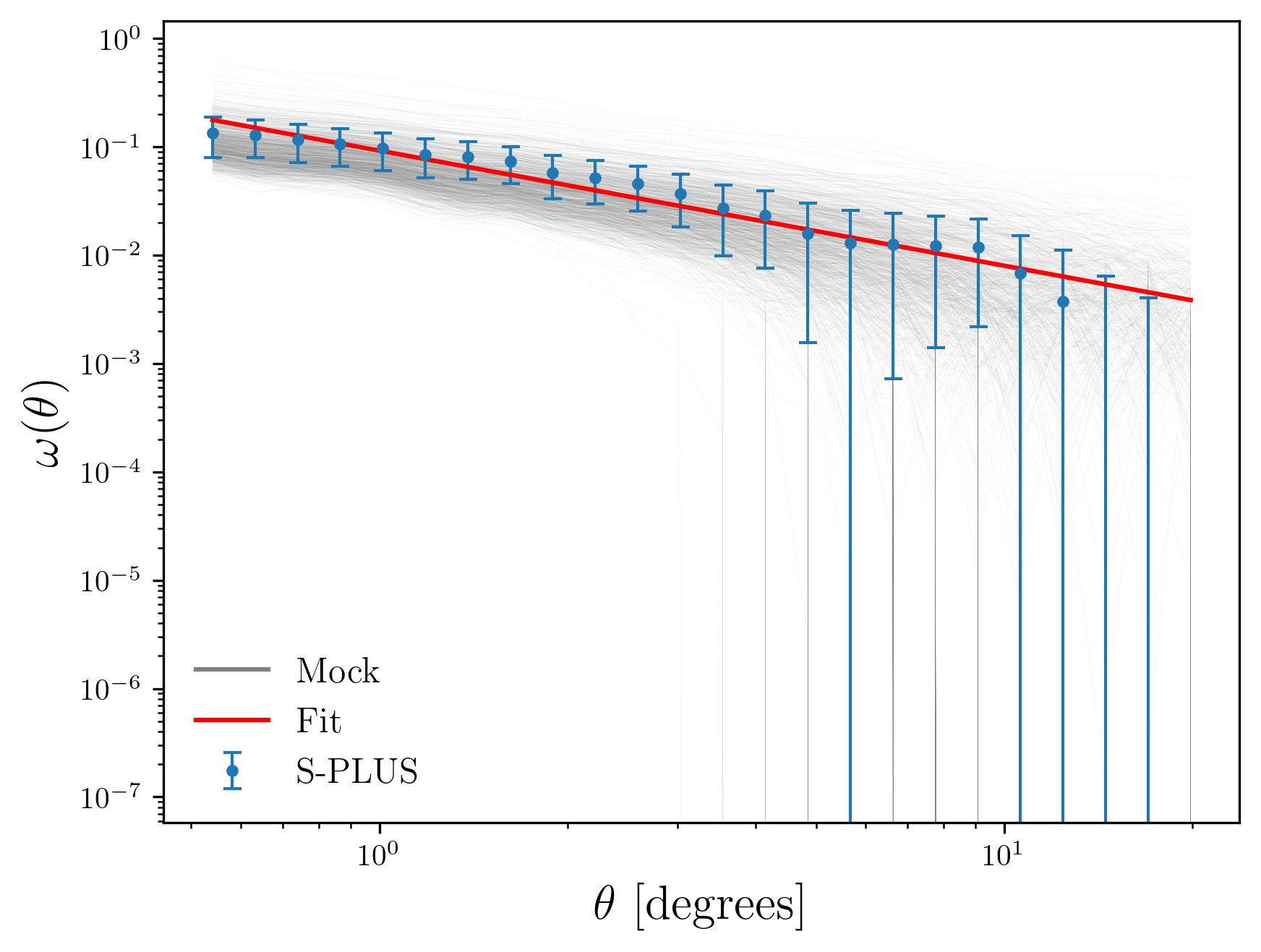}
\caption{2PACF, $\omega(\theta)$, calculated from the blue galaxies sample following the Landy-Szalay methodology (see Sec.~\ref{sec:ls} for details). 
The blue dots represent the binned 2PACF data with $1\sigma$ uncertainties, and the gray curves are the 2PACF computed for each one of the mocks. 
The solid red line corresponds to the best-fit model (see Eq.~(\ref{eq:2pacf-power})), with $\beta = 0.11 \pm 0.04$ and $\theta_0 = 1.06 \pm 0.13\;{\rm deg}$.} 
\label{fig:tpacf_ls}
\end{figure}

Our first analyses are done with the Landy-Szalay estimator, described in Sec.~\ref{sec:ls}. 
For this, we calculated the 2PACF for our data sample within the angular interval $[0.5^{\circ}, 20^{\circ}]$, as presented in Fig.~\ref{fig:tpacf_ls}. 
The blue markers are obtained from the data, while the ensemble of gray lines shows $1,000$ mock realizations generated with the same sample geometry. 
The power-law trend is well captured by the fitting of Eq.~\eqref{eq:2pacf-power}, outlined in the red, continuos line. 
Note that, even at the smallest angles, the GLASS simulations reproduce the observed clustering amplitude, indicating that our mocks are suitable reproductions of the data.

\begin{figure*}[!ht]
\centering
\includegraphics[width=0.45\linewidth]{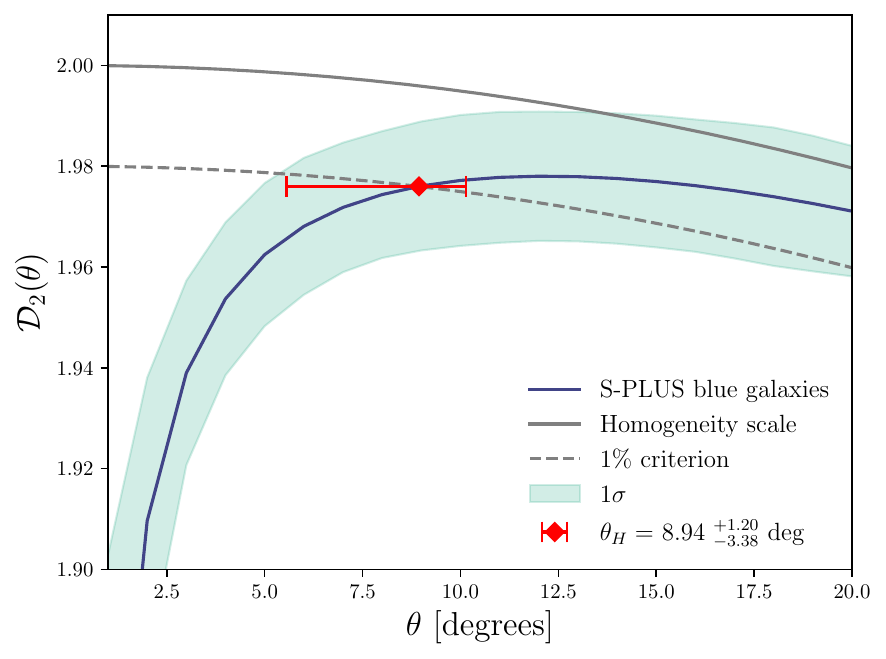}
\includegraphics[width=0.45\linewidth]{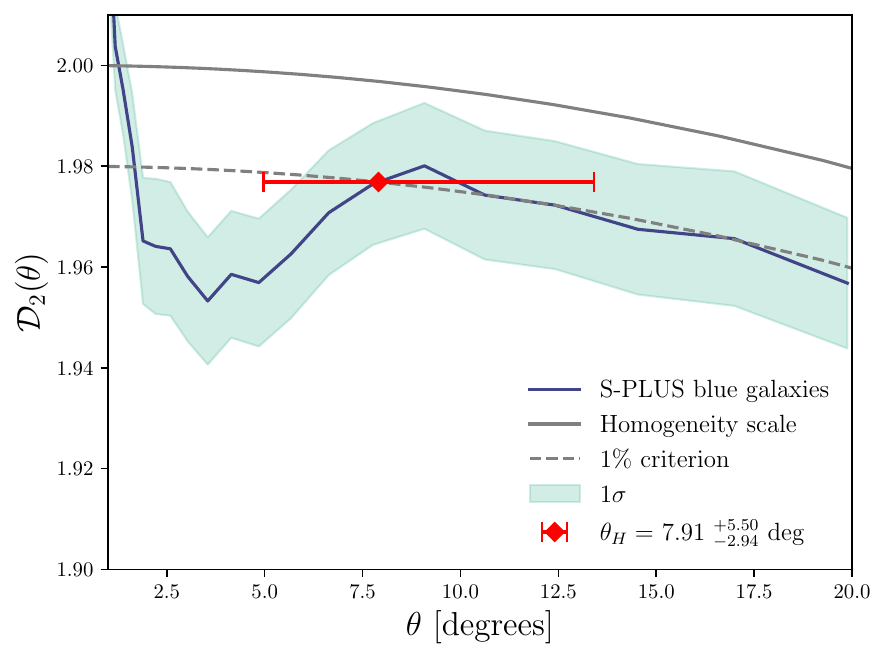}

\caption{Correlation dimension, $\mathcal{D}_2(\theta)$. 
Left: LS methodology. The solid blue curve was obtained from Eq.~\eqref{eq:d2_2d}, with the shaded region representing the $1\sigma$ interval. The solid gray line marks the theoretical expectation for a perfectly homogeneous angular distribution, while the dashed gray line corresponds to the $1\%$ criterion. 
The red diamond indicates the angular homogeneity scale, $\theta_H = 8.94_{-3.38}^{+1.20}\;{\rm deg}$, defined as the transition where the measured $\mathcal{D}_2(\theta)$ meets the criterion. 
Right: The same, but for the AFCD methodology using Eq.~\eqref{eq:d2_omegas}. The angular scale of homogeneity in this case is $\theta_H = 7.91_{-2.94}^{+5.50}\;{\rm deg}$. 
The behavior observed at very small angles is a purely numerical artifact. 
Because the right side of Eq.~(\ref{eq:wb}) diverges in the limit 
$\theta \rightarrow 0$, the numerical evaluation of the function $\bar{\omega}(\theta)$ at 
$\theta = 0$ generates unrealistic large values of $\mathcal{D}_{2}(\theta)$ for $\theta \approx 0$.
}
\label{fig:d2_ls_smooth}
\end{figure*}

The measured $\mathcal{D}_2(\theta)$ over the range $\theta \in [1^{\circ}, 20^{\circ}]$ is presented in Fig.~\ref{fig:d2_ls_smooth}. 
Applying the $1\%$ criterion of Eq.~\eqref{eq:1percent}, we find the angular scale of homogeneity 
\begin{equation*}
\label{thetaH1}
    \theta_H = 8.94_{-3.38}^{+1.20}\;{\rm deg} \,,
\end{equation*}
as summarized in Table~\ref{tab:theta_h}.

\begin{table*}[!ht]
\centering
\caption{Angular scale of transition to homogeneity, $\theta_{H}$, and the respective $\Lambda{\rm CDM}$ predictions, for both methodologies.}
\begin{tabularx}{0.9\linewidth}{>{\centering\arraybackslash}X>{\centering\arraybackslash}X >{\centering\arraybackslash}X >{\centering\arraybackslash}X>{\centering\arraybackslash}X>{\centering\arraybackslash}X>{\centering\arraybackslash}X}
\hline
\hline
& \multicolumn{3}{c}{LS estimator} & \multicolumn{3}{c}{AFCD estimator}\\
\cmidrule(lr){2-4} \cmidrule(lr){5-7} 
& Data & Mocks & $\Lambda$CDM & Data & Mocks & $\Lambda$CDM\\
\midrule
$\theta_{H}$ & $8.94_{-3.38}^{+1.20}\;{\rm deg}$ & $10.17\;{\rm deg}$ & $8.14\, {\rm deg}$ & $7.91_{-2.94}^{+5.50}\;{\rm deg}$ & $5.65\;{\rm deg}$ & $8.09\, {\rm deg}$\\
\bottomrule
\end{tabularx}
\label{tab:theta_h}
\end{table*}

\begin{figure*}[!ht]
\centering
\includegraphics[width=0.45\linewidth]{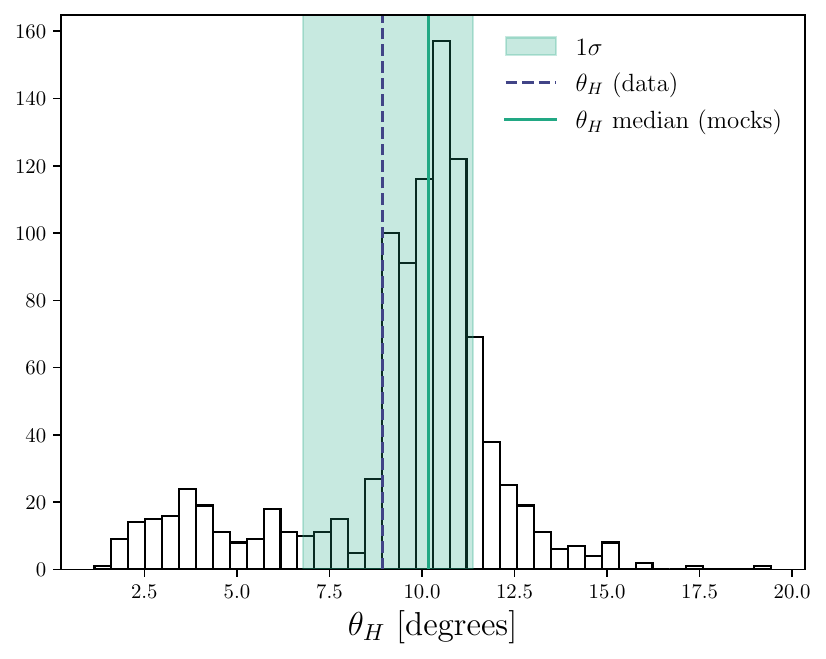}
\includegraphics[width=0.45\linewidth]{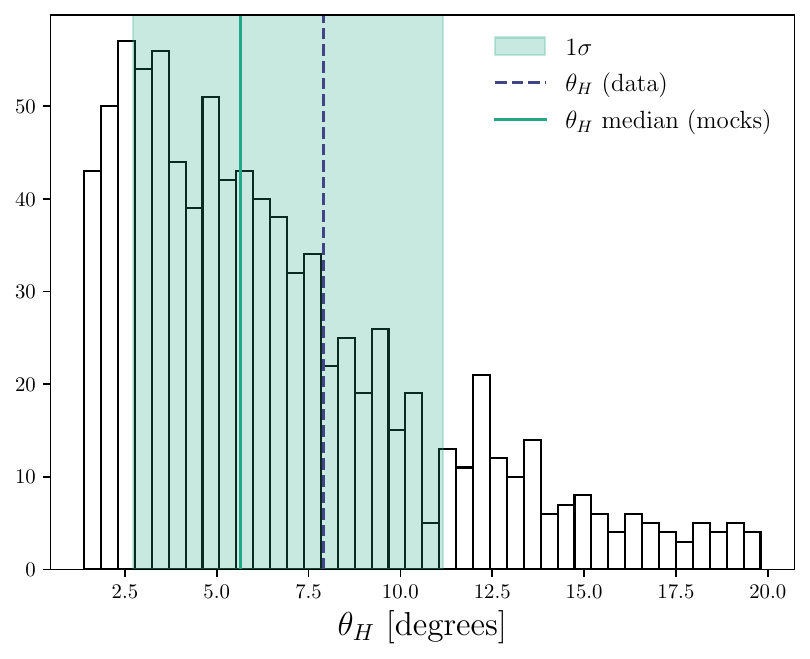}

\caption{Distribution of angular homogeneity scales measured from $1,000$ mock catalogs (see Table~\ref{tab:theta_h} for a summary of 
our results). 
Left: The LS methodology (see Sec.~\ref{sec:ls} for details). 
The green vertical line indicates the median $\tilde{\theta}_H = 10.17\,{\rm deg}$, obtained from the mocks, with the green shaded region corresponding to the $1\sigma$ confidence interval. 
The blue dashed line corresponds to the value obtained analysing the S-PLUS blue galaxies sample, that is, $\theta_H = 8.94\,{\rm deg}$. 
Right: The AFCD methodology  (see Sec.~\ref{sec:d2_2d} for details). 
The green vertical line indicates the median value $\tilde{\theta}_{H} = 5.65\,{\rm deg}$, obtained from the mocks, with the green shaded region corresponding to the $1\sigma$ confidence interval. 
The blue dashed line corresponds to the S-PLUS data measurement $\theta_H = 7.91\,{\rm deg}$.}
\label{fig:hist_ls_smooth}
\end{figure*}

To assess the statistical robustness, we measured $\theta_H$ in $1,000$ mocks generated with GLASS. 
Figure~\ref{fig:hist_ls_smooth} displays the distribution of the values across these mocks.

The green vertical line corresponds to the median value, $\tilde{\theta}_H = 10.17\,{\rm deg}$ for the set of $1,000$ mocks, and the green shaded region is the $1\sigma$ interval. Therefore, the deviation is within $0.4\sigma$ confidence interval.

%%------------------------------------------------
\subsection{Results with AFCD methodology}
As presented in Section~\ref{sec:d2_2d}, the result obtained via Eq.~\eqref{eq:d2_omegas} differs from the previous one in that it employs a more direct approach, without the need to fit a power-law model, and that relies exclusively on the profile of $\omega(\theta)$. 
Therefore, $\mathcal{D}_2(\theta)$ is calculated directly from the data, being a valid approximation of the classical counts-in-caps method. 

We employ this approach by computing $\omega(\theta)$ on $24$ logarithmically spaced bins for $\theta \in [0.5^{\circ}, 20^{\circ}]$
and evaluating the correlation dimension using Eq.~\eqref{eq:d2_omegas}. Figure~\ref{fig:d2_ls_smooth} shows the resulting $\mathcal{D}_2(\theta)$. 

Applying the $1\%$ criterion of Eq.~\eqref{eq:1percent}, we found that the resulting angular scale of homogeneity is 
\begin{equation*}
    \theta_H = 7.91_{-2.94}^{+5.50}\;{\rm deg}.
\end{equation*}
% at the effective redshift $z_{\rm eff} = 0.12$. 
% As we did previously, this value corresponds to $\sim 39\,h^{-1}\;{\rm Mpc}$.

The histogram of $\{ \theta_H \}$ values obtained from the mocks is shown in Fig.~\ref{fig:hist_ls_smooth}. 
The deviation from the mocks median $\tilde{\theta}_{H} = 5.65\,{\rm deg}$ value is within $0.4\sigma$ CL.

The nonparametric AFCD approach differs from the previous one in the way the fractal dimension, $\mathcal{D}_{2}(\theta)$, is calculated. In the parametric approach, $\mathcal{D}_{2}(\theta)$ is derived from $\mathcal{N}(<\theta)$, which reduces the impact of statistical fluctuations at large angular scales. 
However, this attenuation feature tends to shift the transition scale 
to slightly larger angles. 
In contrast, the nonparametric approach, instead of deriving 
$\mathcal{N}(<\theta)$, directly applies Eq.~\eqref{eq:d2_omegas}, 
which avoids numerical differentiation of the correlation function. 
This is particularly advantageous, since numerical derivatives are more 
sensitive to statistical noise, binning effects in the 2PACF, and 
rounding errors than numerical integration. 
As a consequence, the method reduces the impact of statistical fluctuations at 
large angular scales, though it tends to anticipate the transition when small 
fluctuations push $\mathcal{D}_2(\theta)$ to the threshold value.

%%--------------------------------------------------
\subsection{Theoretical prediction of \texorpdfstring{$\theta_{H}$}{}}
\label{sec:theta_theoretical}

To quantify the theoretical expectation of the angular transition to cosmic homogeneity, we compute the 2PACF predicted by the flat-$\Lambda$CDM cosmology. 
This quantity encapsulates, in projection, the statistical imprint of the three-dimensional matter distribution on the celestial sphere, and is given by the double integral over the redshift~\citep{Crocce2011},
\begin{equation}
    \omega_{\rm th}(\theta) = \int_{0}^{\infty}{dz_{1}\,\phi(z_{1}}) \int_{0}^{\infty}{dz_{2}\,\phi(z_{2})\,\xi(s)}\;,  
\end{equation}
where $\xi(s)$ is the spatial two-point correlation function evaluated at the comoving separation $s$ and $\phi(z)$ is the normalized selection function describing the redshift distribution of galaxies, $n(z)$, as 
\begin{equation}
    \phi(z) = \frac{n(z)}{\int{n(z)dz}}\;.
\end{equation}

Given two cosmic objects at redshifts $z_{1}$ and $z_{2}$, respectively, separated by and angle $\theta$, the comoving separation considering a Friedmann-Lemaître-Robertson-Walker 
metric is 
\begin{equation}
s = \sqrt{\chi^{2}(z_{1}) + \chi^{2}(z_{2}) - 2\chi(z_{1})\chi(z_{2})\cos{\theta}} \,,
\end{equation}
where $\chi(z)$ is the comoving radial distance 
\begin{equation}
\chi (z) = \frac{c}{H_{0}} \int_{0}^{z}{\frac{dz^{\prime}}{\sqrt{\Omega_{m}(1 + z)^{3} + \Omega_{\Lambda}}}} \,,
\end{equation}
for the flat-$\Lambda$CDM model, which we calculate using the Core Cosmology Library~\footnote{\url{https://github.com/LSSTDESC/CCL}}~\citep[\textsc{CCL}; ][]{Chisari19}.

The spatial two-point correlation function is related to the matter power-spectrum, $P(k, z)$, through the Fourier transform
\begin{equation}
\xi(s,z) = \int_{0}^{\infty}{\frac{dk}{2\pi^{2}}\,k^{2}\,
j_{0}(ks)\,P(k,z)} \,,
\end{equation}
where $j_{0}$ is the spherical Bessel function of the first kind of order zero. 

To compute $P(k, z)$, we use the \textsc{CAMB}\footnote{\url{https://camb.info/}} code~\citep{camb}, adopting a flat-$\Lambda$CDM cosmology with parameters consistent with the last \cite{Planck2018}, as summarized in Table~\ref{tab:table_mock}.

After obtaining the theoretical correlation function, we applied the same methodologies described in Secs.~\ref{sec:ls}~and~\ref{sec:d2_2d} to determine the theoretical angular scale of transition to homogeneity. The resulting values are $\theta_{H}^{\Lambda{\rm CDM}} = 8.14\,{\rm deg}$ for the LS method and $\theta_{H}^{\Lambda{\rm CDM}} = 8.09\,{\rm deg}$ for the AFCD method. These estimates are summarized in Table~\ref{tab:theta_h}, alongside the observational measurements obtained from the data. 

We assess the agreement with $\Lambda$CDM by computing the percentile rank~\citep{Wasserman04} of the observed LS and AFCD values within the mock distribution. 
The LS method lies at the $22$th percentile of the mocks ($p$ value = $0.45$), while the AFCD lies at the $69$th percentile ($p$ value = $0.61$). 
This means that our results from both approaches are 
consistent with each other and in good agreement with the scales predicted by the standard cosmological model $\Lambda$CDM. 
This conclusion is supported by the comparison analysis done in the 
next subsection.

%%------------------------------
\subsection{Comparison with previous measurements}
\label{sec:comparison}

\begin{figure}[!ht]
    \centering
    \includegraphics[width=0.9\linewidth]{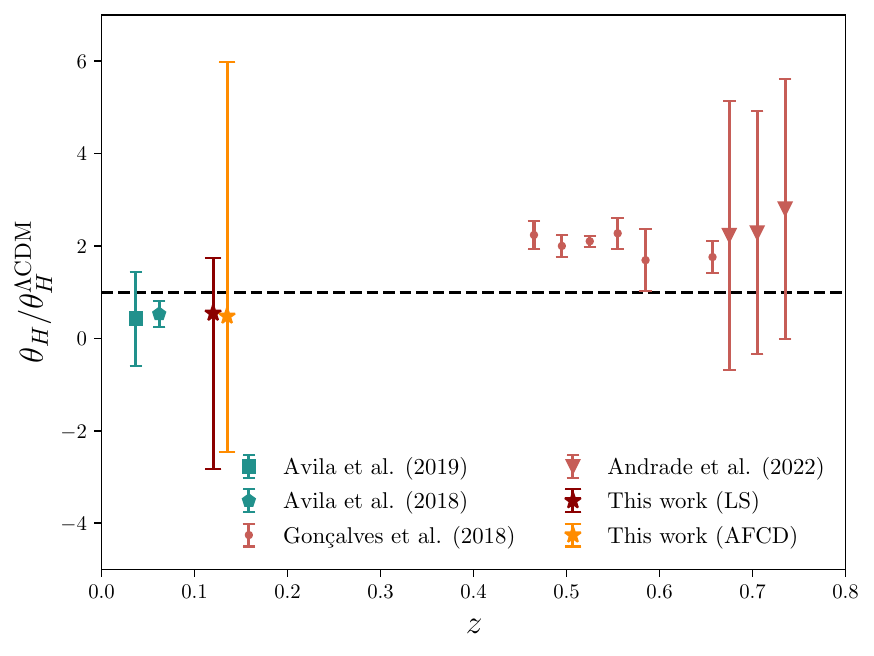}
    \caption{Comparison between our estimates of the angular homogeneity scale and previous measurements from the literature, obtained with different tracers and methodologies~\citep{Avila2018, Avila2019,Goncalves2018,Andrade2022}. The dark red and orange stars correspond to the LS and AFCD estimates obtained in this work, respectively. The literature measurements are color-coded according to the tracer used in each analysis: blue symbols correspond to studies based on blue galaxies or HI sample (which are expected to have galaxy bias close to unity; see Appendix~\ref{app:bias}), while red symbols indicate analyses based on luminous red galaxies (which typically present bias $b > 1$.) 
    % The dashed black line shows the expected evolution of the angular homogeneity scale in the $\Lambda$CDM framework, assuming a constant comoving homogeneity scale.
    The dashed black line corresponds to the perfect agreement between the observed scale and the prediction from the $\Lambda$CDM model.}
    \label{fig:thetah-comp}
\end{figure}

To place our results in context, we compare our estimates with previous measurements available in the literature, obtained with diverse tracers and methodologies. Figure~\ref{fig:thetah-comp} shows that our LS and AFCD measurements are consistent with them across different redshift ranges.

Because the value of $\theta_H$ depends on the tracer~\citep{Scrimgeour2012}, the overall redshift dependence follows the expected behaviour in the $\Lambda$CDM framework. 
Our results, therefore, complement previous model-independent analyses by providing a new estimate using blue galaxies from the S-PLUS survey in the local Universe. 
Therefore, we conclude that our results, from both approaches, are in good agreement with the angular scale of homogeneity predicted by the standard cosmological model $\Lambda$CDM.

%%------------------------------
\section{Robustness tests via resampling}
\label{sec:robustness}

Photometric uncertainties can significantly impact tomographic analyses, altering the number of galaxy pairs within a given redshift slice and, thereby, biasing the inferred angular correlation function. To mitigate these effects and to ensure the reliability of our results, we implement a resampling technique~\citep{Ribeiro2025,Asorey2016} based on the individual redshift probability distribution function (PDF) provided for each galaxy in the S-PLUS catalog. 
Each PDF, $p_{i}(z)$, is modeled as a weighted sum of Gaussian components, 
\begin{equation}
p_{i}(z) = \sum_{k}{w_{ik}\,\mathcal{N}(z|\mu_{ik}, \sigma_{ik})},
\end{equation}
where $w_{ik}$, $\mu_{ik}$, and $\sigma_{ik}$ denote the weight, mean and standard 
deviation of the $k$th Gaussian component, respectively. 
These parameters capture the multimodal and asymmetric structure of photometric redshift estimates~\citep{Herpich2024}\footnote{For more information about the columns in the data catalog, see, for example, \url{https://datalab.noirlab.edu/data-explorer?showTable=splus_dr4.photoz} and \url{https://www.splus.cloud/documentation/DR4} and the references therein.}. 

We employ the inverse transform sampling method to generate realizations of the galaxy distribution consistent with these PDFs. For each galaxy $i$, we first construct its normalized PDF on a redshift grid $z \in [0, 1]$ with $10,000$ sampling points. The cumulative distribution function 
\begin{equation}
C_{i}(z) = \int_{z_{min}}^{z}{p_{i}(z^{\prime})dz^{\prime}},
\end{equation}
is then computed. 
Random values $u_{j} \in [0, 1]$ are drawn from a uniform distribution, and the corresponding sampled redshifts are obtained as $z_{ij} = C_{i}^{-1}(u_{j})$, 
with $i$ and $j$ denoting the $i$th galaxy and the $j$th resampling realization, respectively. 
This procedure is repeated for all galaxies to produce one complete realization of the dataset. 
Repeating the process $N_d = 1,000$ times yields a set of resampled catalogs, each representing a statistically consistent dataset of the observed data driven by the photometric-redshift uncertainties.

\begin{figure}[!ht]
    \centering
    \includegraphics[width=0.9\linewidth]{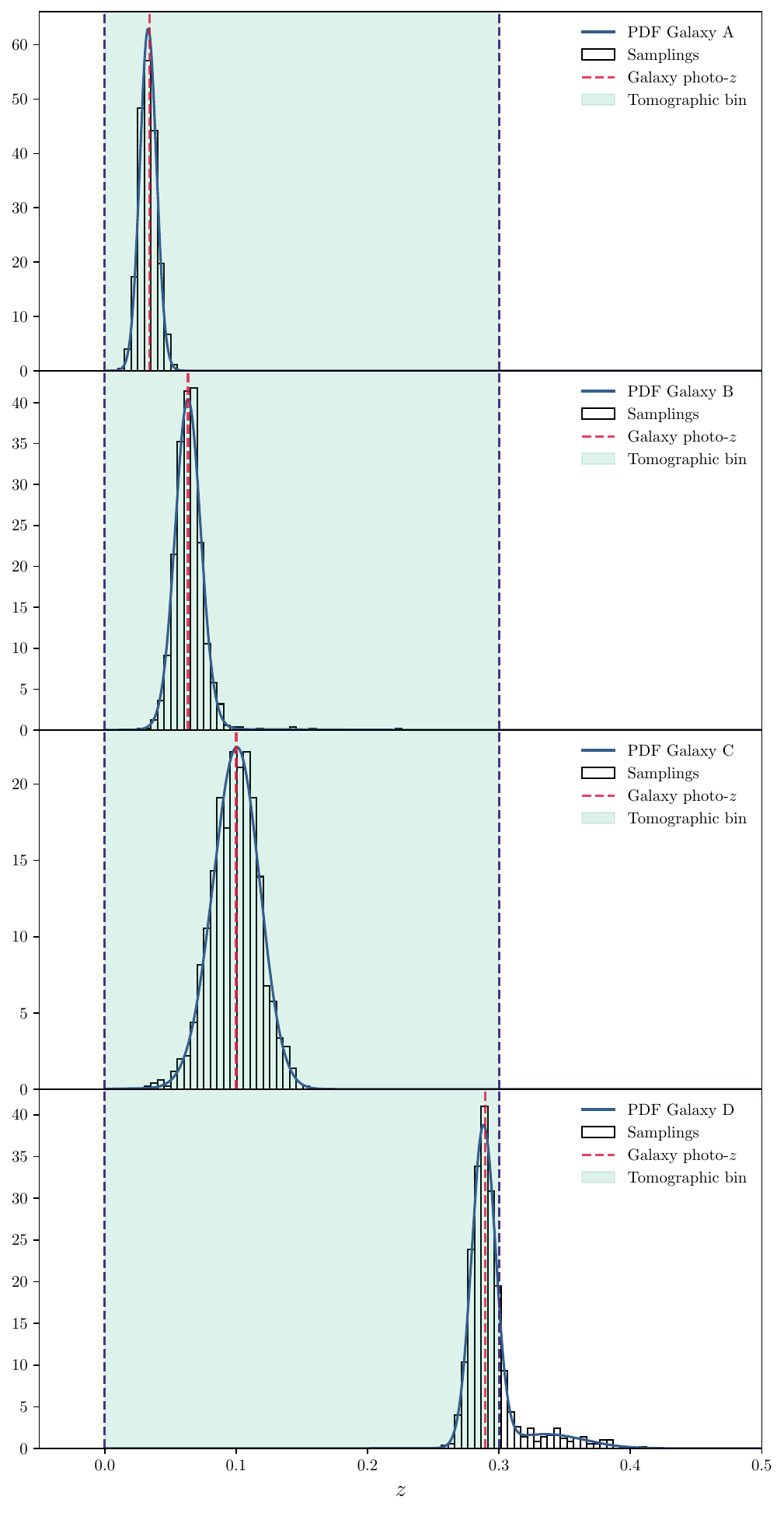}
    \caption{Sampling of galaxies. For each galaxy, the solid curve represents its reconstructed redshift PDF, $p(z)$, while the histogram shows the distribution of $10,000$ redshift draws obtained through random resampling of the Gaussian mixture components using the weights. 
    The red dashed lines mark the photometric redshift from the catalog, and the shaded region corresponds to the tomographic slice adopted in the analysis.}
    \label{fig:resampling}
\end{figure}

Each realization is then subject to the same tomographic selection used in the analysis. Specifically, galaxies are assigned to redshift slices of width $\Delta z$, chosen to be larger than the typical redshift uncertainty, $\langle \sigma_{z} \rangle$, in order to minimize boundary effects. In this way, the analysis slice covers $z \in [0, 0.3]$, while the PDFs are defined on a broader grid $z \in [0, 1]$, ensuring that the tails of the distributions are fully represented. In this sense, galaxies whose PDFs partially overlap the tomographic boundaries can contribute to multiple realizations, thus preserving the statistical information encoded in the photo-$z$ uncertainties. Figure~\ref{fig:resampling} illustrates the resampling method for a subset of galaxies.

\begin{figure}[!ht]
\centering
\includegraphics[width=0.9\linewidth]{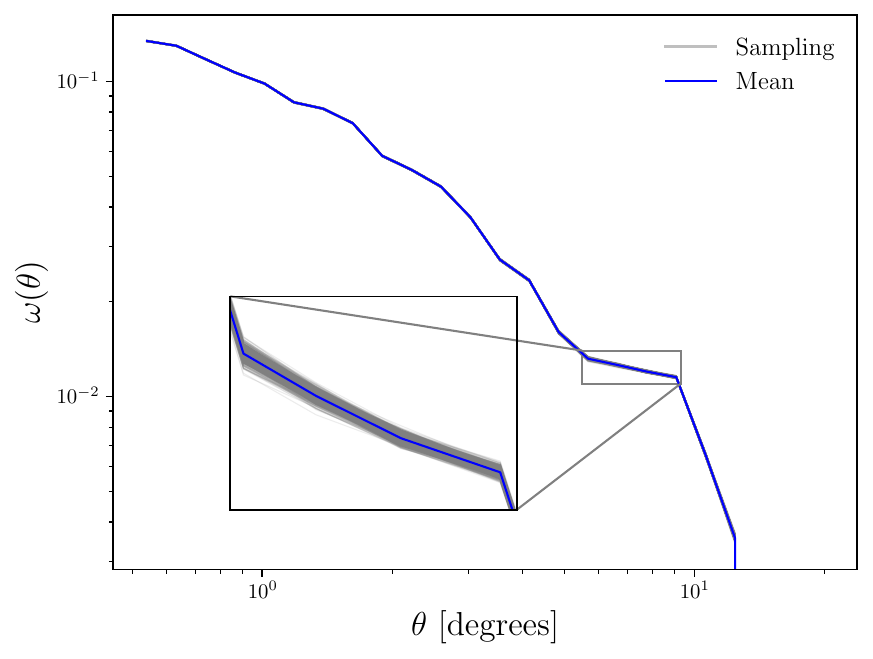}
\caption{2PACF computed for the set of $1,000$ resampled catalogs. 
The gray lines correspond to the 2PACF for each individual sampling, while the blue line denotes their mean. 
The horizontal axis corresponds to that in the Fig.~\ref{fig:tpacf_ls},
% ~and~\ref{fig:tpacf_smooth}, respectively, 
shown in Sec.~\ref{sec:results}. 
% \textbf{Upper panel:}  Analyses done on the resampled catalogues using the LS methodology. 
% \textbf{Bottom panel:} Analyses done on the resampled catalogues using the AFCD methodology.
The inset highlights the small deviation between the curves.}
\label{fig:resampling-tpacf}
\end{figure}

The ensemble of $N_d = 1,000$ resampled catalogs produced an equal number of 2PACF measurements that are statistically consistent with each another, exhibiting only a small dispersion around the mean correlation function, as can be seen in Fig.~\ref{fig:resampling-tpacf}. 
The modest spread can be directly attributed to the combination of a relatively broad tomographic bin, i.e. $z \in [0, 0.3]$ and the narrowness of the individual photo-$z$ PDFs. Because the vast majority of galaxies have $p_{i}(z)$ sharply peaked well inside the adopted interval, their inclusion or exclusion in any given realization is virtually unchanged. Consequently, the effective angular footprint of the sample remains stable and the 2PACF, being purely angular, shows minimal sensitivity to the perturbations in $z$.

This outcome can be interpreted as follows: the measured clustering signal is robust against uncertainties in $z$, 
and the resampling analysis confirms that the angular correlation measurements can be safely used. It is worth noting, however, that the features observed in this resampling is specific to the chosen redshift interval. For narrower tomographic bins or broader $p(z)$ distributions, the same procedure would yield noticeably larger dispersion, as more galaxies would cross the bin boundaries between realizations (see, for example, \cite{Ribeiro2025}).

%%------------------------------
\section{Conclusions}
\label{sec:conclusions}

The observed universe is highly inhomogeneous, in particular in the local Universe, with huge matter structures and large voids distributed everywhere~\citep{Einasto1991, Franco2025a, Courtois13, Courtois2025, Novaes2025, Gavas2025, Telles2000}. 
On the other hand, inhomogeneities were tiny in the distant past, as revealed by the CMB temperature fluctuations, therefore there was a cosmic evolution from homogeneously distributed density fluctuations to a clumpy inhomogeneous universe~\citep{Marques20, Huterer2023}. 
This leads to the natural question regarding a 
minimum scale, at any epoch, from which one can consider the matter content in the universe as homogeneously distributed~\citep{Alonso2014, Avila2022}. 

Thus, in this work, we presented a model-independent estimate of the angular homogeneity scale in the local Universe using $27,289$ blue galaxies within the redshift range $0 \leq z < 0.3$ from the S-PLUS catalog. 
We applied two complementary estimators --a parametric power-law fit to the 2PACF and a nonparametric fractal correlation dimension analysis-- obtaining consistent results for the homogeneity scale, with $\theta_H = 8.94_{-3.38}^{+1.20}\;{\rm deg}$ and $\theta_H = 7.91_{-2.94}^{+5.50}\;{\rm deg}$, respectively.
Some words are due regarding the differences observed in the fractal dimension function $\mathcal{D}_{2}(\theta)$ shown in Fig.~\ref{fig:d2_ls_smooth}. 
In fact, while the LS methodology smooths the 2PACF through the power law 
given in Eq.~(\ref{eq:2pacf-power}), the AFCD approach uses the angular 
correlation function $\omega(\theta)$ computed from the S-PLUS catalog, 
where clusters of galaxies and large voids are present (making the 2PACF noisy). 
The difference in the way $\mathcal{D}_{2}(\theta)$ is calculated is reflected in the behavior observed in these figures. 
A second point deserving a comment concerns the large uncertainties of our $\theta_H$ estimates, in both approaches, which is mainly related to the size of the redshift bin, $z \in [0, 0.3]$. 
This bin spans a large volume of the universe, and the large uncertainties in $\theta_H$ indicate that the estimated scale is valid for the diverse epochs of the universe covered by this redshift bin.

Given that our galaxy sample consists of blue galaxies, with a bias $b \simeq 1$, our estimates can be directly compared with the angular homogeneity scale for matter ($b = 1$) predicted by the $\Lambda$CDM model.  
Moreover, the measurements obtained with the LS and AFCD approaches are also in good agreement with the results obtained analyzing the set of mock catalogs following both methodologies. 

Tests of homogeneity require a sufficiently wide sky coverage and a suitable number density of cosmic objects, as the transition occurs on large scales. As stated in Sec.~\ref{sec:splus}, the analyzed S-PLUS region covers $1,696\,{\rm deg}^{2}$. 
While this area already allows a meaningful test of the homogeneity scale in the local Universe, a natural extension of this work includes applying 
the same methodology to a wider area and, consequently, more populated future data release of S-PLUS. 
Nevertheless, from the current results, we can already conclude that  the blue galaxies sample from the S-PLUS has the potential to validate the CP regarding the angular scale of transition to homogeneity, providing a measurement of such scale, $\theta_H$, in the local Universe.

This work was carried out using computational resources provided by the Sci-Mind servers machines developed by the CBPF AI LAB team and the Data Processing Center of the National Observatory (CPDON).

%------------------------------
\section*{Acknowledgments}
C. F. thanks Coordenação de Aperfeiçoamento de Pessoal de Nível Superior (CAPES) for the financial support. 
F. A. thanks Fundação de Amparo à Pesquisa do Estado do Rio de Janeiro (FAPERJ), Processo SEI-260003/001221/2025, for the financial support.
A. B. acknowledges a CNPq fellowship.

%------------------------------
\section*{Data Availability}
The data that support the findings of this article are not publicly available. 
The data are available from the authors upon reasonable request. 
%------------------------------
\bibliography{bib}  

%------------------------------
\appendix
%------------------------------
\section{Robustness test of the random catalog density}
\label{app:random}

\begin{figure}[!ht]
\centering
\includegraphics[width=0.9\linewidth]{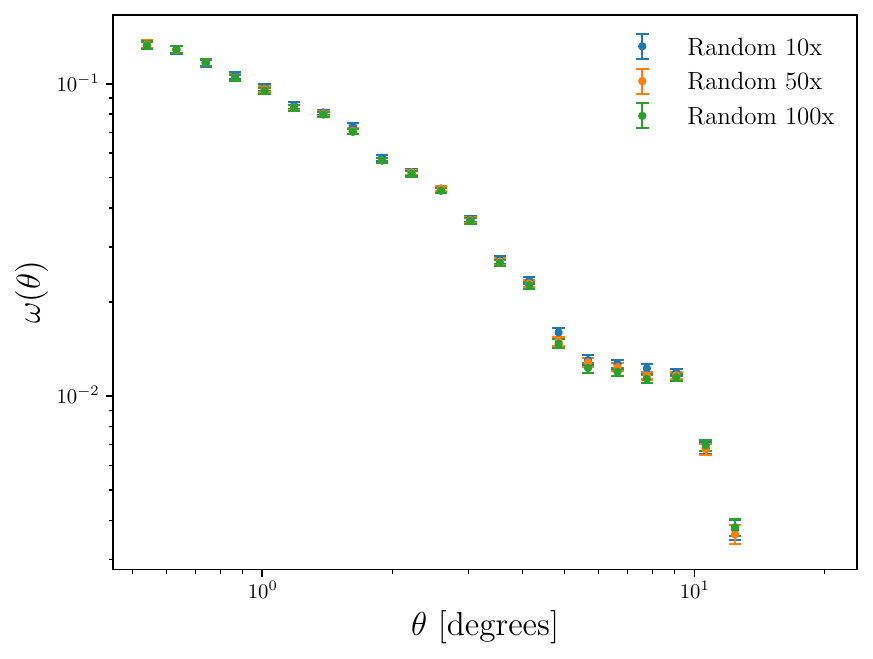}
\caption{Comparison of the results obtained with random catalogs ten (blue), 50 (orange), and 100 (green) times denser than the data.}
\label{fig:tpacf-comp}
\end{figure}

To assess the impact of the random catalog number density in our measurements, we have performed a comparative analysis using random catalogs containing ten, 50, and 100 times the number of data points. Figure~\ref{fig:tpacf-comp} shows the resulting 2PACF for each case, with the uncertainties corresponding to the internal ones estimated by \textsc{TreeCorr}.

We find that the results are stable against changes in the number density of the random catalog. 
Therefore, we conclude that considering a random catalog with ten times the number of data points is sufficient for the purposes of our analyses.

%------------------------------
\section{Deduction of Eq.~\texorpdfstring{\eqref{eq:d2_omegas}}{}}
\label{appendixA}

The cumulative scaled counts-in-caps within an angular separation $\theta$ is defined as
\begin{equation}
\label{eq:scaled1}
    \mathcal{N}(<\theta) \equiv 1 + \bar{\omega} (\theta) \,,
\end{equation}
where $\bar{\omega}(\theta)$ corresponds to the average angular correlation function, given by
\begin{equation}
\label{eq:bar-omega}
    \bar{\omega}(\theta) \equiv \frac{1}{1 - \cos{\theta}}\int_{0}^{\theta}{\omega(\theta^\prime)\sin{\theta^\prime d\theta^\prime}} \,.
\end{equation}
Using these relationships in Eq.~\eqref{eq:d2_2d}, one can derive the expression for the correlation dimension, as follows 
\begin{equation}\label{eq:corr_dim_rw}
\begin{split}
\mathcal{D}_{2}(\theta) & = \frac{d\ln{\mathcal{N}(<\theta)}}{d\ln{\theta}} + \frac{\theta \sin{\theta}}{1 - \cos{\theta}}\\
& = \frac{\theta}{\mathcal{N}(<\theta)} \frac{d\mathcal{N}(<\theta)}{d\theta} + \frac{\theta \sin{\theta}}{1 - \cos{\theta}}\\
& = \frac{\theta}{1 + \bar{\omega}(\theta)} \frac{d\bar{\omega}(\theta)}{d\theta} + \frac{\theta \sin{\theta}}{1 - \cos{\theta}} \,.
\end{split}
\end{equation}
For convenience, let us introduce the notation
\begin{equation}
I(\theta) \equiv \int_{0}^{\theta}\omega(\theta^\prime)\sin{\theta^\prime d\theta^\prime},
\end{equation}
then, the derivative of the averaged correlation function can be 
expressed as
\begin{equation}
\label{eq:dw_dtheta}
\frac{d\bar{\omega}}{d\theta} = \frac{\sin{\theta}}{(1 - \cos{\theta})^{2}}[\omega(\theta)(1 - \cos{\theta}) - I(\theta)]\,.
\end{equation}
Finally, substituting this result into Eq.~\eqref{eq:corr_dim_rw}, 
the correlation dimension simplifies to 
\begin{equation}
\mathcal{D}_2(\theta) = \frac{\theta \sin{\theta}}{1 - \cos{\theta}}\left[\frac{1 + \omega(\theta)}{1 + \bar{\omega}(\theta)}\right] \,, 
\end{equation}
which is the Eq.~\eqref{eq:d2_omegas}. 
This final expression provides a direct relation between the correlation dimension and the 2PACF, highlighting how departures from homogeneity can be quantified at a given angular scale. \\

%%------------------------------
\section{Sensitivity to galaxy bias}
\label{app:bias}
A consistency check of the galaxy bias was performed by comparing the measured 2PACF with the theoretical matter angular correlation function computed from the $\Lambda$CDM matter power spectrum. Assuming the linear relation~\citep{Fry1993}
\begin{equation}
    \omega (\theta) = b^{2}\,\omega_{th}(\theta),
\end{equation}
we estimate the bias in the range $\theta \in [1^{\circ}, 20^{\circ}]$. We find a value of $b \simeq 1.06$, supporting the approximation adopted in this work.

\begin{figure}[!ht]
    \centering
    \includegraphics[width=0.9\linewidth]{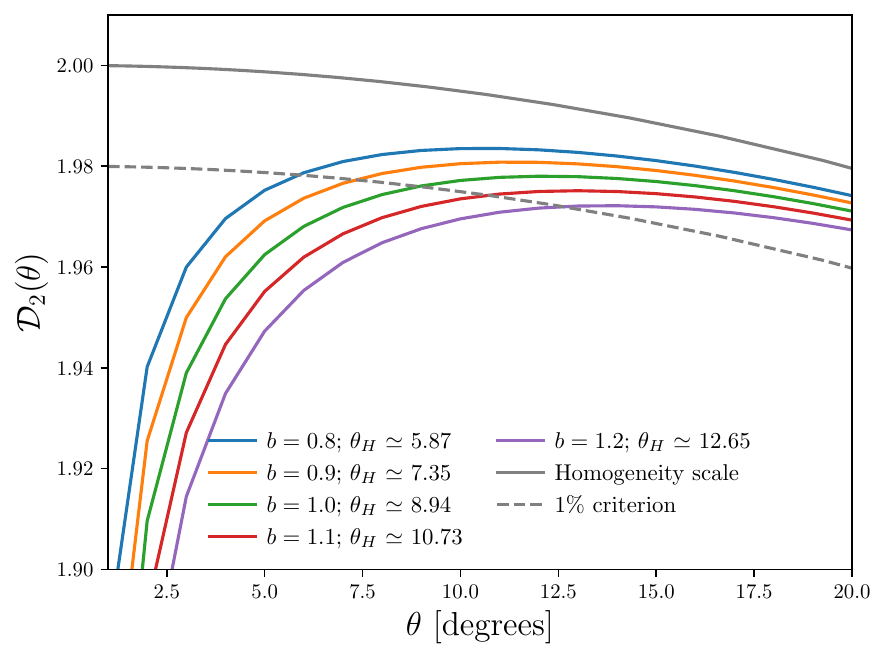}
    \includegraphics[width=0.9\linewidth]{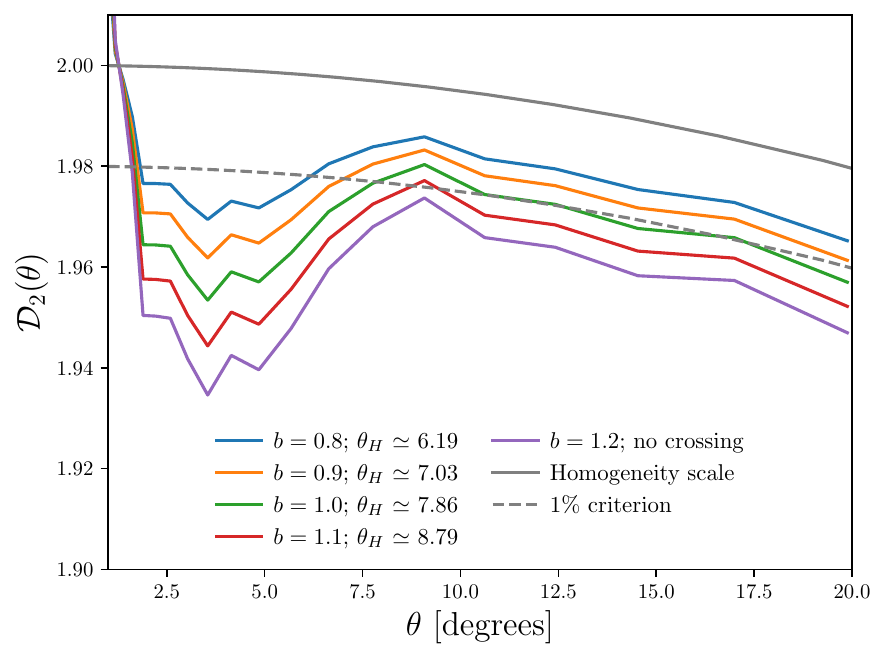}
\caption{Correlation dimension computed assuming different galaxy bias values, $b = \{ 0.8, 0.9, 1.0, 1.1, 1.2 \}$. 
The upper and lower panels show the analyses for the LS and AFCD methodologies, respectively. 
As expected, the transition scale  to homogeneity is increased with higher bias values.}
\label{fig:bias_d2}
\end{figure}

Furthermore, to evaluate the sensitivity of the inferred angular homogeneity scale to the assumed galaxy bias, we computed $\theta_H$ considering several values for the bias, namely 
$b = \{ 0.8, 0.9, 1.0, 1.1, 1.2 \}$. 
The resulting $\mathcal{D}_{2}(\theta)$ functions and the corresponding estimates for $\theta_H$, for the LS and AFCD methodologies, are shown in Fig.~\ref{fig:bias_d2}. 

As expected, larger (smaller) bias values shift the homogeneity scale toward larger (smaller) angles. 

\end{document}